\documentclass{vak2024}
\usepackage[utf8]{inputenc}
\usepackage{url}
\usepackage[pdfpagelabels=false]{hyperref}	
\hypersetup{colorlinks=true,linkcolor=blue,citecolor=blue,filecolor=blue,urlcolor=blue,}
\newcommand{\micron}{~$\mu m$ }
\newcommand{\kms}{~km\,s$^{-1}$ }

\begin{document}
\title{Study of the physical and chemical properties of dense clumps in several high-mass star-forming regions}
\titlerunning{Dense clumps in HMSFRs}  
\author{
    A. G. Pazukhin\inst{1,2}
    \and
    I. I. Zinchenko\inst{1}
    \and
    E. A. Trofimova\inst{1}}
\authorrunning{Pazukhin et al.} 
	%
	%
\institute{A.V. Gaponov-Grekhov Institute of Applied Physics of the Russian Academy of Sciences, 46 Ul’yanov str., 603950 Nizhny Novgorod, Russia
    \and 
    Lobachevsky State University of Nizhny Novgorod, 23 Gagarin Ave, 603950 Nizhny Novgorod, Russia}
\abstract{
 {Massive stars play an important role in the Universe. Unlike low-mass stars, the formation of these objects located at great distances is still unclear. It is expected to be governed by some combination of self-gravity, turbulence, and magnetic fields.}
   {Our aim is to study of the chemical and physical conditions of dense clumps in several high-mass star-forming regions.}
   {We performed observations towards 5 high-mass star-forming regions (L1287, S187, S231, DR21(OH), NGC7538) with the IRAM 30 m telescope. We covered the 2-3 and 4 mm wavelength band and analyzed the lines of HCN, HNC, HCO$^+$, HC$_3$N, HNCO, OCS, CS, SiO, SO$_2$ and SO. Using astrodendro algorithm on the 850\micron dust emission data from the SCUBA Legacy catalogue, we identified dense gas clumps and determined their masses, H$_2$ column densities and sizes. Furthermore, the kinetic temperatures, molecular abundances and dynamical state were obtained. The Red Midcourse Space Experiment Source survey (RMS) was used to determine the clump types.}
   {We identified $\sim$20 clumps. We found no significant correlation between line width and size, but the linewidth-mass and mass-size relationships are strongly correlated. Virial analysis indicated that the clumps with HII regions and young stellar objects (YSOs) are gravitationally bound. Furthermore, it was suggested that significant magnetic fields provide additional support for clump stability. The molecular abundances decrease from YSOs to submm and HII regions.}
\doi{10.26119/VAK2024-ZZZZ}
}

\maketitle

\section{Introduction}

High-mass stars (also OB stars, $>10^3\, L_\odot$, $>8\, M_\odot$) play an important role in the Universe via their feedback. Their formation and evolution are still unclear.  According to the review by \citet{Zinnecker07}, the evolution phase can be divided into several groups of objects. Objects associated with the first phase of high-mass star formation have been called IR dark clouds (IRDCs), having dense and cold gas likely represent the initial conditions of high-mass star formation. Hot molecular cores (HMCs) have large masses of warm and dense gas, large abundances of complex organic and maser emission. Finally, ionized radiation from the embedded stellar population disrupts the parent molecular cloud, and an ultra-compact HII region (UCHII) is formed.
\section{Observational data}\label{obs}
In September 2019, with the IRAM 30 m radio telescope, we observed five massive star forming regions (in the framework of the project 041-19). The list of sources is given in Table~\ref{tab:source}. Observations were carried out in the on-the-fly (OTF) mode over a mapping area of $200''\times200''$ and covered the 2-3 and 4 mm wavelength band. A more detailed description of the observations and data reduction can be found in our published paper on deuterated molecules~\citep{Pazukhin23}. The following lines of molecules were observed: HCN~(1-0), HNC~(1-0), HCO$^+$~(1-0) and their $^{13}$C isotopologues, HC$_3$N~(8-7), OCS~(6-5), C$^{34}$S~(3-2), SiO~(2-1), SO~(2-1), SO$_2$~(6$_{0,6}$-5$_{1,5}$) and HNCO~(4$_{0,4}$-3$_{0,3}$).
\begin{table}[h]
\centering
\caption{List of sources.} \label{tab:source}
\begin{tabular}{lcccccccccccccccccccc}
 \hline
            Source & RA(J2000) & Dec(J2000) & V$_{\rm lsr}$ & d & R$_{\rm GC}$  \\
                   & h:m:s & d:m:s & \kms & kpc & kpc   \\
            \hline
            L1287     & 00:36:47.5 & 63:29:02.1 & -17.7 & 0.93 & 8.64  \\
            S187     & 01:23:15.4 & 61:49:43.1 & -14.0 & 1.44 & 9.06 \\
            S231     & 05:39:12.9 & 35:45:54.0 & -16.6 & 1.56 & 9.67 \\
            DR21(OH) & 20:39:00.6 & 42:22:48.9 &  -3.8 & 1.50 & 8.04\\
            NGC7538  & 23:13:44.7 & 61:28:09.7 & -57.6 & 2.65 & 9.48\\
            \hline
\end{tabular}       
\end{table}

\section{Results}
To extract the clumps, we use Python astrodendro\footnote{\url{http://www.dendrograms.org}}~\citep{Rosolowsky08}. A dendrogram is employed to represent a hierarchical data structure. The dendrogram consists of two types of structures: branches, which can be divided into branches and leaves, which represent the final structures. We define the leaves on the dendrogram as clumps. 
We identified $\sim$20 clumps. The RMS survey~\citep{Lumsden13} was used to determine the clump types.  Three clumps were found to be associated with the HII regions, 10 with YSOs, and 7 with submillimetre emission.
 Masers were identified from the maser database\footnote{\url{https://maserdb.net}}~\citep{Ladeyschikov19}. Water, hydroxyl, and methanol masers are observed in clumps associated with YSO and HII regions.
 
Using 850\micron dust emission data from the SCUBA Legacy catalogue~\citep{scuba}, we determined the masses, H$_2$ column density and size of the clumps. To estimate the clump masses, the following equation was used~\citep{Kauffmann08}:
\begin{equation}
  \begin{array}{l}
    M = 
    \displaystyle 0.12 \, M_{\odot}
    \left( {\rm e}^{14.39 (\lambda / {\rm mm})^{-1}
        (T / {\rm K})^{-1}} - 1 \right) 
     \left( \frac{\kappa_{\nu}}{\rm cm^2  g^{-1}} \right)^{-1}\\
    \quad \displaystyle
    \cdot \left( \frac{F_{\nu}}{\rm Jy} \right)
    \left( \frac{d}{\rm 100~pc} \right)^2
    \left( \frac{\lambda}{\rm mm} \right)^{3}
  \end{array}
\end{equation}
where the gas-to-dust ratio is assumed to be 100,
$\lambda$~is the wavelength,
$F_{\rm\nu}$~is the integrated flux, 
$T_{\rm dust}$~is the dust temperature,
$d$~is the distance to the object.
Dust opacity $\kappa_{\rm \nu}=1.82$~cm$^2$g$^{-1}$ at 850~µm~\citep{Ossenkopf94}.
The dust temperature $T_{\rm dust}$ was assumed to be 20~K.  
Thus, the H$_2$ column density and volume density are derived as $N(H_2) = \frac{M}{\mu_{\rm H_{2}} m_{\rm  H} (\pi R_{\rm eff}^2)}$ and $n(H_2) = \frac{M}{\mu_{\rm H_{2}} m_{\rm  H} (4/3\pi R_{\rm eff}^3)},$
where $m_{\rm H}$~ is the hydrogen mass, the mean molecular weight $\mu_{\rm H_2}=2.8$, $R_{\rm eff}$ is the effective radius determined from the clump area $\sqrt{A/\pi}$. 
 
The virial parameter, which compares the mass of the virial with the gas mass, provides one method to investigate the cloud stability.  The virial parameter is defined as follows~\citep{Bertoldi92}:
$\alpha_{\rm vir}=\frac{5\sigma_{\rm tot}^2R_{\rm eff}}{GM},$
where $G$~ is the gravitational constant. The $\sigma_{\rm tot}$ is defined by the width of the observed line considering the thermal and non-thermal component.

 We adopted the kinetic temperature maps derived from the integral intensities ratios of the $J=1-0$ HCN and HNC and their $^{13}$C isotopologues from~\citet{Pazukhin23}.
 The column density is estimated assuming local thermodynamic equilibrium conditions, the optically thin, Rayleigh-Jeans and negligible background approximation.
 
\begin{figure}[t]
\begin{minipage}{0.333\linewidth}
        \centerline{\includegraphics[width=.99\linewidth]{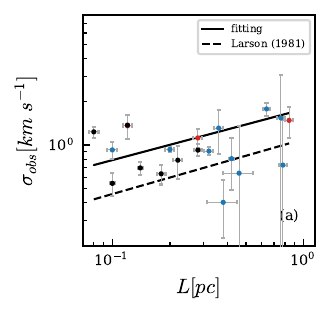}}
    \end{minipage}\hfill   
\begin{minipage}{0.333\linewidth}
        \centerline{\includegraphics[width=.99\linewidth]{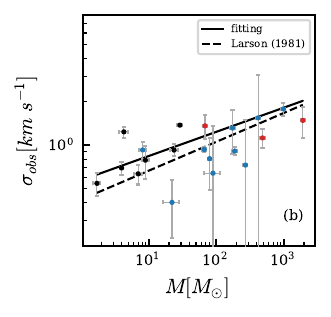}}
    \end{minipage}\hfill
\begin{minipage}{0.333\linewidth}
        \centerline{\includegraphics[width=.99\linewidth]{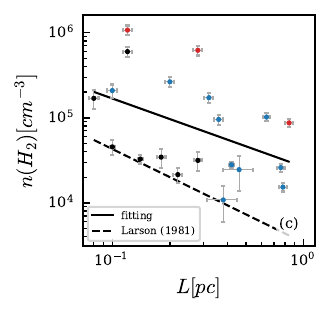}}
    \end{minipage}\vfill   
\begin{minipage}{0.333\linewidth}
        \centerline{\includegraphics[width=.99\linewidth]{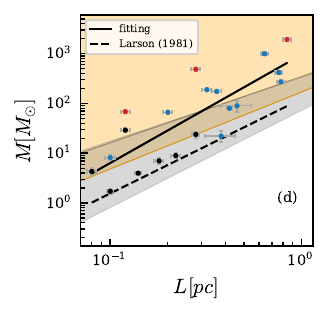}}
    \end{minipage}
\begin{minipage}{0.333\linewidth}
        \centerline{\includegraphics[width=.99\linewidth]{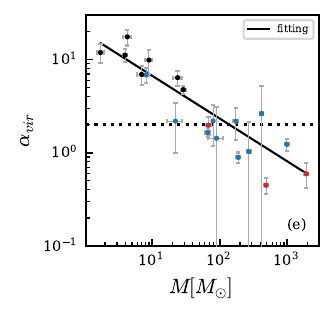}}
    \end{minipage}    
\caption{The relations of (a) $\sigma_{\rm obs}-L$, (b) $\sigma_{\rm obs}-M$, (c) $n({\rm H_2})-L$, (d) $M-L$ and (e) $\alpha_{\rm vir}-M$ for clumps associated with submm emissions~(black), YSOs~(blue) and HII regions~(red). In panel~(d), the orange shading indicates the region in which magnetic field is necessary for the clump stability against gravitational collapse. Region without massive star formation highlighted by gray shading \citep{Kauffmann13}. In all panels, the black line shows the fitting result. The black dashed line demonstrates the original Larson relation~\citep{Larson81}.}
\label{fig:params_cor}
\end{figure}

\citet{Larson81} defined the relations between linewidth, cloud mass, volume density, and cloud size. The Larson's laws are as follows: (1) the velocity dispersion as a function of the cloud size, $\sigma \propto L^{0.38}$, (2)  the velocity dispersion as a function of the cloud mass, $\sigma \propto M^{0.2}$, and (3) the mean density as a function of the cloud size, $n \propto L^{-1.1}$. Then, the mass-size relationships can be defined from (1) and (2), $M \propto L^{1.9}$.

In Figure~\ref{fig:params_cor} we show the relation of $\sigma_{\rm obs}-L$, $n({\rm H_2})-L$, $M-L$ and $\sigma_{\rm obs}-M$.
The results indicate a weak correlation between the $\sigma_{\rm obs}$ and $L$, as evidenced by the Spearman's rank correlation coefficient $r_{\rm s}=0.15$ (p-value=0.533). We estimate the power-law index to be $0.35\pm0.15$. 
The correlation between $\sigma_{\rm obs}$ and $M$ is strong ($r_{\rm s}=0.6$, p-value=0.008), with a slope of $0.17\pm0.05$. 
The $n(\rm H_2)$ and $L$ show a strong anti-correlation with $r_{\rm s}=-0.5$ (p-value=0.023) and a slope of $-0.81\pm0.38$.
The mass and size present a strong correlation ($r_{\rm s}=0.8$, p-value=$2.4\times10^{-5}$), with a power-law index of $2.2\pm0.37$.

Figure~\ref{fig:params_cor}(d) demonstrates mass-size relationships and magnetic fields. The clumps with YSO and HII regions are situated within the magnetic field region ($B>0\, \mu G$), and exhibit a low virial parameter ($\alpha_{\rm vir}<2$). The density and mass of these clouds is such that the thermal pressure and random gas motions are insufficient to provide significant support against self-gravity. This is evidenced by a low virial parameter, which suggests that magnetic fields provide additional support.

\begin{figure}[t]
\centerline{\includegraphics[width=.6\textwidth]{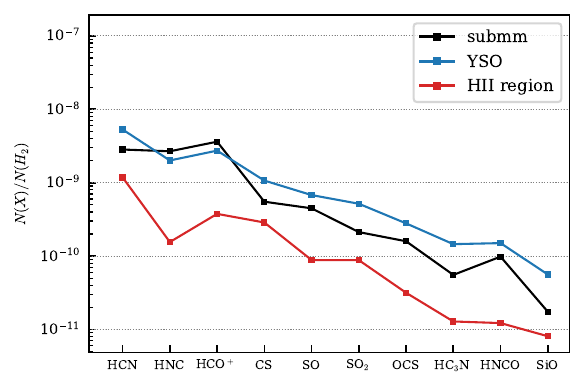}}
\caption{The mean abundances relative to H$_2$ for clumps associated with submm emissions~(black), YSOs~(blue) and HII regions~(red).}
\label{fig:abun}
\end{figure}

Figure~\ref{fig:abun} illustrates histograms of molecular abundances for different clump types. \citet{Yu16} have investigated 87 RMS sources (28 MYSOs and 59 HII regions).
Two subgroups of MYSOs and HII regions have been found no significant difference. \citet{Gerner14} observed a sample of 59 sources and modeled the chemical evolution including different evolutionary stages. The abundances have been increased with evolutionary phase. More complex and heavy molecules are formed with evolving age until the HMC phase and decline for the UCHII stage,  when its are by the UV-radiation from the embedded sources. The molecular abundances agree with our results, assuming uncertainties of about 10. 

\section{Summary}
In this work, we study the physical and chemical conditions of five high-mass star-forming regions, using observations with the IRAM-30m radio telescope. The results are as follows:
\begin{enumerate}

\item We identified $\sim$20 clumps. Three clumps were found to be associated with the HII regions, 10 with YSOs, and 7 with submillimetre emission. The clumps have typical sizes of $\sim$0.2~pc and masses $\sim$100$-1000\, M_\odot$, kinetic temperatures of $\sim$20$-40\, K$ and line widths of $\sim$2\kms. 

\item We found no significant correlation between line widths and sizes, but the linewidth-mass and mass-size relationships are strongly correlated. Virial analysis indicated that the clumps with HII regions and YSOs are gravitationally bound. Virial parameter dependence on the mass of $\alpha_{\rm vir}\propto M ^{-0.5}$. Furthermore, it was suggested that significant magnetic fields provide additional support for clump stability. 

\item The molecular abundances decrease from clumps with YSOs to submm and HII regions. The maximum abundance was obtained for HCN ($\sim$10$^{-9}$), and decreased to $\sim$10$^{-11}$ for SiO.
\end{enumerate}

\acknowledgements{
The research is based on observations made by the 041-19 project with the 30-m telescope. IRAM is supported by INSU/CNRS (France), MPG (Germany) and IGN (Spain). This paper also made use of information from the RMS survey data base at \url{http://rms.leeds.ac.uk/} which was constructed with support from the Science and Technology Facilities Council of the UK.}

\section*{Funding} 
This study was supported by the Russian Science Foundation grant No. 24-12-00153.


\end{document}